\documentclass[prl,amsmath,aps,showpacs,twocolumn,superscriptaddress]{revtex4}
\usepackage{graphicx,xspace}
\usepackage{float}
\usepackage{amsmath}
\usepackage{amssymb}

\input{epsf}

\usepackage{dcolumn}
\usepackage{bm}

\begin{document}

\title{Topological defects and misfit strain in magnetic stripe domains of lateral multilayers with perpendicular magnetic anisotropy}

\author{A. Hierro-Rodriguez}
\affiliation{Depto. F{\'i}sica, Universidad de Oviedo, 33007
Oviedo, Spain} \affiliation{CINN, (CSIC-UO-P. Asturias), Llanera,
Spain}

\author{R. Cid}
\affiliation{Depto. F{\'i}sica, Universidad de Oviedo, 33007
Oviedo, Spain} \affiliation{CINN, (CSIC-UO-P. Asturias), Llanera,
Spain}

\author{M. V\'elez}
\email{mvelez@uniovi.es} \affiliation{Depto. F{\'i}sica,
Universidad de Oviedo, 33007 Oviedo, Spain}

\author{G. Rodriguez-Rodriguez}
\affiliation{Depto. F{\'i}sica, Universidad de Oviedo, 33007
Oviedo, Spain}  \affiliation{CINN, (CSIC-UO-P. Asturias), Llanera,
Spain}

\author{J. I. Mart{\'i}n}
\affiliation{Depto. F{\'i}sica, Universidad de Oviedo, 33007
Oviedo, Spain} \affiliation{CINN, (CSIC-UO-P. Asturias), Llanera,
Spain}

\author{L. M. \'Alvarez-Prado}
\affiliation{Depto. F{\'i}sica, Universidad de Oviedo, 33007
Oviedo, Spain}  \affiliation{CINN, (CSIC-UO-P. Asturias), Llanera,
Spain}

\author{J. M. Alameda}
\affiliation{Depto. F{\'i}sica, Universidad de Oviedo, 33007
Oviedo, Spain} \affiliation{CINN, (CSIC-UO-P. Asturias), Llanera,
Spain}

\begin{abstract}
Stripe domains are studied in  perpendicular magnetic anisotropy
films nanostructured with a periodic thickness modulation that
induces the lateral modulation of both stripe periods and in-plane
magnetization. The resulting system is the 2D equivalent of a
strained superlattice with properties controlled by interfacial
misfit strain within the magnetic stripe structure and shape
anisotropy. This allows us to observe, experimentally for the
first time, the continuous structural transformation of a grain
boundary in this 2D magnetic crystal in the whole angular range.
The magnetization reversal process can be tailored through the
effect of misfit strain due to the coupling between disclinations
in the magnetic stripe pattern and domain walls in the in-plane
magnetization configuration.

\end{abstract}

\maketitle

Stripe domains in ferromagnetic films with perpendicular magnetic
anisotropy (PMA) present a fascinating variety of configurations
ranging from ordered parallel states to disordered labyrinthine
patterns that depend both on material parameters and magnetic and
thermal history \cite{stripes1,stripes2}. They share a common
phenomenology with many other systems with self-organized patterns
such as microdomains in block copolymer thin films
\cite{copolymer}, wrinkles in elastic membranes \cite{wrinkles} or
liquid crystals \cite{liquid crystal}. The physics of stripe
domains is a basic issue both to understand magnetic hysteresis
loops in PMA materials \cite{minor,loop} for technology
applications and to unravel the different phase transitions that
appear in 2D \cite{nelson,saratz}. Studies performed in extended
systems have revealed the complex phase diagram of these modulated
phases \cite{stripes1,clarke,saratz} and the important role of
topological defects in order-disorder mechanisms
\cite{copolymer,venus}. The actual pattern realized in a given
extended sample depends on the interplay between the equilibrium
periodic configuration and the strain present in the magnetic
system and is controlled by the motion of topological defects such
as dislocations, disclinations and grain boundaries
\cite{asciutto,huang}. On the other hand, magnetic stripes in
nanostructured systems, such as dots, rings and wires of PMA
materials \cite{hehn,navas,lee} or copolymers nucleated in
periodic gratings \cite{sundrani} show much simpler patterns due
to the coupling between shape and domain structure \cite{clarke}.

Recently, the concept of magnetic lateral multilayer, i.e. an
extended film with a laterally nanostructured magnetic property
such as anisotropy \cite{ingleses}, saturation magnetization
\cite{McCord} or exchange bias \cite{theis}, has emerged as a
bridge between extended and confined geometries. These laterally
nanostructured samples combine confinement effects and coupling
effects between nearby elements similar to those found in the more
standard geometry of vertical multilayers. In this framework,
magnetic stripe domains in an extended PMA film can be considered
as a bulk 2D crystal of lattice parameter $\lambda$. Then, a
lateral periodic modulation of $\lambda$ would result in the 2D
equivalent of 3D strained superlattices fabricated by the
alternate deposition of layers with different lattice constants
\cite{strained,strained2}. Thus, in the same way as homogeneous
and random strains on the magnetic stripe pattern of extended
samples have a significant influence in their effective
magnetization configuration, the presence of localized strain at
the interfaces of a lateral magnetic stripe multilayer can provide
an extra control over the magnetic hysteresis loop.

In this work, we study the magnetic stripe domain structure in
lateral multilayers fabricated on nanostructured PMA films with
periodic thickness modulation. This has allowed us to observe
experimentally, for the first time, the continuous transformation
of a grain boundary in this 2D magnetic crystal in the whole
angular range driven by misfit strain and, also, how the coupling
between topological defects in the magnetic stripe structure and
the underlying in-plane magnetization configuration can be used to
tailor the magnetization reversal process.

Amorphous 80 nm NdCo$_5$ alloy films have been grown by sputtering
on 10 nm Al/Si(100) substrates, and protected from oxidation with
a 3 nm Al capping layer \cite{cid}. At room temperature, the
saturation magnetization is $M_S = 1100$ emu/cm$^3$ and the PMA
$K_N$ is of the order of $10^6$ erg/cm$^3$
\cite{cid,supplemental}. They have been characterized by
Transverse Magnetooptical Kerr effect (TMOKE) with the field $H$
applied parallel to the sample plane and by Magnetic Force
Microscopy (MFM) using a Nanotec$^{TM}$ system with a 1 kOe
electromagnet to apply an in-plane variable $H$ \cite{fernando}.
\begin{figure}[ht]
 \begin{center}
\includegraphics[angle=0,width=1.2\linewidth]{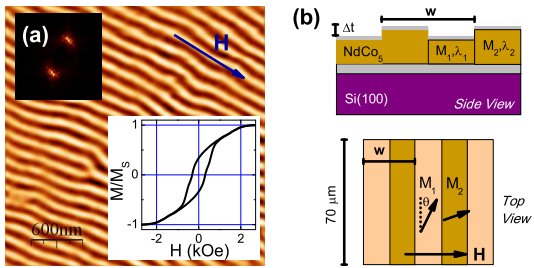}
 \end{center}
 \caption{(color online) (a) MFM image of
 stripe domains in a 80 nm thick Nd-Co film at $H=0$ ($\lambda = 157$ nm).
 Top inset is the FFT pattern. Bottom
  inset is the in-plane TMOKE hysteresis loop.
 (b) Sketch of the nanostructured Nd-Co samples with periodic thickness
 modulation. $\theta$ is the angle measured
 relative to the nanostructured lines.  } \label{Fabricacion}
 \end{figure}
 Figure 1(a) is a MFM image
taken at remanence after applying $H$ = 1 kOe that displays a well
defined stripe domain structure, aligned along the direction of
the last saturating field. The Fast Fourier Transform (FFT) of
this image (top inset of Fig. 1(a)) displays two symmetric peaks
that provide a precise information about the angular orientation
of the stripe pattern and its periodicity ($\lambda = 157$ nm).
The in-plane hysteresis loop (bottom inset of Fig. 1(a)) is a
typical \textsl{transcritical} loop with a linear reversible
region at high fields, characteristic of PMA materials. The finite
value of the remanent magnetization $0.4 M_S$ indicates that,
besides the oscillating out-of-plane magnetization component that
gives rise to the black-white MFM contrast, there is a significant
average in-plane magnetization component $M_{parallel}$, lying
along the stripe domain direction
\cite{luis,clarke,hubert,supplemental}.

For a given set of parameters ($M_S$, $K_N$, $H$), both $\lambda$
and $M_{parallel}$ are a function of sample thickness $t$
\cite{loop,thickness,field}. Thus, a nanostructured sample
composed of alternate linear regions of thickness $t_1$ and $t_2$,
as sketched in Fig. 1(b), would also present a similar lateral
modulation in $\lambda$ and $M_{parallel}$ that is the aim of our
work. In the following we will refer to the stripe period and
in-plane magnetization component in the thin and thick regions as
$\lambda_1$, $M_1$ and $\lambda_2$, $M_2$, respectively. A two
step lithography process has been performed for sample
fabrication. First, $70 \times 70 \mu$m$^2$ flat squares of 80 nm
thick Nd-Co film have been defined by a combined e-beam
lithography and lift-off process. Then, a mask of equispaced
parallel 70 $\mu$m long 10 nm thick Nb lines is defined on top of
the squares by a second combined e-beam lithography and lift-off
process. This pattern of lines is transferred to the underlying
Nd-Co film by ion beam etching with Ar$^+$ ions, creating a set of
linear grooves of depth $\Delta t$ controlled by etching time.
Finally, the sample is covered by a 3 nm Al capping layer.
\begin{figure}[ht]
 \begin{center}
 \includegraphics[angle=0,width=0.8\linewidth]{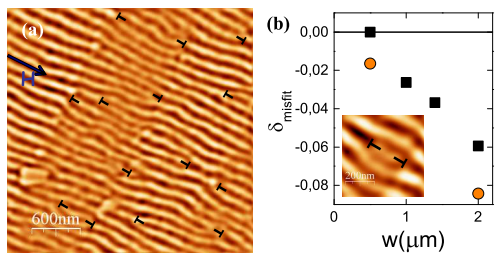}
 \end{center}
 \caption{(color online)
 (a) $3 \times 3 \mu$m$^2$ MFM image of
 stripe domains in nanostructured Nd-Co film ($\Delta t = 12$ nm, $w = 2 \mu$m) at $H=1$ kOe.
 (b) Misfit strain \textit{vs.}
 lateral periodicity at $H = 1$ kOe: ($\circ$), $\theta \simeq 90^\circ$;
 ($\Box$), $\theta \simeq 5^\circ$. Inset is a detail of misfit dislocations
 in the sample with $w = 0.5 \mu$m.}
 \end{figure}The result is a film with alternate linear regions of
thickness $t_1 = 80$ nm $- \Delta t$ and $t_2 = 80$ nm, width
$w/2$ and lateral period $w$. Two series of samples have been
fabricated either with shallow ($\Delta t$ = 12 nm) or deep
($\Delta t$ = 30 nm) grooves  and $w =$ 0.5, 1, 1.4 and 2 $\mu$m,
in order to analyze the behavior of these lateral multilayers in
the two limits of either small or large amplitude thickness
modulation. A flat $70 \mu$m Nd-Co square has also been defined
near each nanostructured sample for control purposes.

 Figure 2(a) shows the MFM image of a Nd-Co sample
 with shallow grooves ($\Delta t = 12$ nm) and $w=2 \mu m$ taken in
 $H = 1$ kOe applied at $\theta \simeq 90^\circ$, showing a well defined magnetic stripe pattern oriented along $H$. The
effect of nanostructuration in their configuration is clear:
 the thinner regions, present weaker and more closely spaced stripes with $\lambda_1 = 132$ nm, whereas
 in the thicker regions the stripes show a stronger contrast and a
 larger period $\lambda_2 = 144$ nm (topography images have been used
 as a mask to select the relevant area for the FFT analysis of stripe periods in
 thin and thick regions).

Several dislocations can be identified within the magnetic stripe
pattern, about half of them located at the edges between
thin-thick lines: two dislocations appear in the left-most edge,
whereas four dislocations can be counted in the central edge. In
all the cases they correspond to the addition of an extra stripe
to the pattern in the thin region, i.e. they can be identified as
misfit dislocations. In the control 80 nm flat film, measured at
the same conditions, the magnetic stripe period is $\lambda_0 =
151$ nm and only 3 dislocations can be seen in a similar $3 \times
3 \mu$m$^2$ area. Thus, two kinds of strains appear in the
magnetic system: first, an effective misfit strain that can be
defined as
\begin{equation}
\delta_{misfit} = \frac{\lambda_1-\lambda_2}{\lambda_2}
\end{equation}
and is relaxed by misfit dislocations; second, a residual strain
due to the difference with the equilibrium magnetic stripe period
that results in elastic energy stored in the system. In this case,
$\delta_{misfit} = - 0.083$ which, taking the Burgers vector $b =
132$ nm, implies an average misfit dislocation spacing
\cite{strained,strained2} $D = b/|\delta_{misfit}|$ = 1580 nm.
This is equivalent to two dislocations in a 3 $\mu$m long edge,
which is qualitatively in agreement with the observed numbers in
Fig. 2(a).

As $w$ decreases, $\lambda$ values in the thin and thick lines
approach to each other and the absolute value of misfit strain
becomes smaller both for $H$ parallel and perpendicular to the
nanostructured lines (see Fig. 2(b)). For example, for $w = 0.5
\mu$m, $\delta_{misfit} = 0$ with $H$ at $\theta = 5^0$ and
$\delta_{misfit} = -0.016$ with $H$ at $\theta = 90^0$. In this
last case, misfit dislocations appear as closely bound pairs
spaced at $w/2$ (see inset of Fig. 2(b)). This is different from
the behavior of isolated PMA wires \cite{lee} in which stripe
period is independent of wire width for $H$ perpendicular to the
edges, remarking the relevance of interaction between magnetic
stripes in neighboring lines in our experiment. Actually, the data
in Fig. 2(b) follow the characteristic trend of strained
superlattices \cite{strained,strained2}, in which
$|\delta_{misfit}|$ is an increasing function of layer thickness
($w/2$ in our case) above a critical thickness given by the
balance between dislocation and elastic energies. The critical
line width here can be estimated as $w_c/2 \approx 250$ nm, which
is of the order of $1.5 \lambda_0$. That is, for smaller feature
sizes misfit strain should be negligible and the magnetic stripe
pattern becomes coherent over the whole sample. It can be noted
that previous works in patterned PMA films were in this small
feature limit (feature size of the order of $\lambda_0$) and,
thus, only domain pinning effects were reported
\cite{pinning,pinning2}.

The magnetization reversal process is almost the same in films
with shallow grooves and in flat films: stripe domains stay
parallel to $H$ during the whole hysteresis loop with a small
enhancement in $\lambda$ at the coercivity, in a similar way as
reported for other PMA films \cite{loop,field}. However, in the
samples with deeper grooves the differences between thin and thick
regions are enhanced and the magnetic behavior changes
qualitatively. Figure 3 shows a series of MFM images of a
nanostructured Nd-Co film ($\Delta t = 30$ nm, $w = 1.4 \mu$m)
taken at increasing fields after saturation at $H=-1$ kOe
perpendicular to the lines \cite{epaps}. In the thin regions, a
well defined pattern of parallel stripes is seen in all the images
that rotates in a continuous fashion away from the applied field
direction and becomes aligned to the nanostructured lines at
coercivity ($H_C = 90$ Oe). On the other hand, stripes in the
thick regions remain always  oriented approximately along the
applied field direction but develop a labyrinthine structure at
coercivity. Thus, a variable angle grain boundary appears at the
interface between thin and thick lines that undergoes a continuous
structural transformation during the magnetization reversal
process.
\begin{figure}[ht]
 \begin{center}
\includegraphics[angle=0,width=0.8\linewidth]{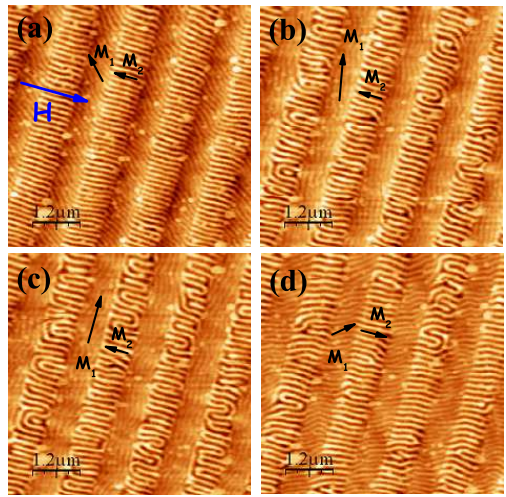}
 \end{center}
 \caption{(color online)
  $6 \times 6 \mu$m$^2$ MFM images of
 stripe domains in nanostructured Nd-Co film ($\Delta t = 30$ nm, $w = 1.4 \mu$m)
 taken after saturation at $H=-1$ kOe at: (a) $H=-12$
 Oe; (b) $H = 74$ Oe; (c) $H=96$ Oe; (d) $H = 150$ Oe. Arrows indicate the positive sense of the applied
 field and the average in-plane magnetization in the thick and thin
 regions.}
 \end{figure}

 The magnetization rotation in the thin regions (see Fig. 4(a)) can be attributed
 to the effective shape anisotropy created by the flux discontinuities
 that appear at the interface between thin and thick lines due to the lateral
 modulation of the in-plane magnetization \cite{lateral}. In this framework, the
 leading energy terms within the thin lines correspond to the
 dipolar and Zeeman terms. Then, the energy density $e$
 for $M_1$ oriented at $\theta$ relative to the lines, $M_2$ at $90^0$ and $H$ at $\theta_0$ may be written as:
\begin{equation}
e = 2\pi N_x (M_1\sin\theta-M_2)^2 - H M_1 \cos(\theta-\theta_0)
\end{equation}
with $N_x$ the demagnetizing factor perpendicular to the lines.
Thus, the equilibrium magnetization orientation would be given by
\begin{equation}
H = -4\pi N_x M_1 \frac{\sin \theta \cos
\theta}{\sin(\theta-\theta_0)}+ 4\pi N_x M_2 \frac{\cos
\theta}{\sin(\theta-\theta_0)}.
\end{equation}

 The first term corresponds to a rotation process under an
anisotropy field $H_K = 4\pi N_x M_1$ and the second to the bias
field created by the thick lines $H_D = 4\pi N_x M_2$ weighed by
an angular factor that is close to unity for $\theta_0 \approx
90^0$. Figure 4(b) is a plot of $\sin \theta \cos
\theta/\sin(\theta-\theta_0)$ \textit{vs.} $H$ with
$\theta_0=95^0$. A linear behavior appears both for $H < 50$ Oe
and $H > 150$ Oe with $H_K = 235$ Oe in both cases and $H_D =
-135$ Oe and $70$ Oe respectively. Since $H_D$ and $M_2$ are
proportional, the change of sign in $H_D$ can be taken as a
signature of magnetization reversal in the thick regions.

\begin{figure}[ht]
 \begin{center}
\includegraphics[angle=0,width=1\linewidth]{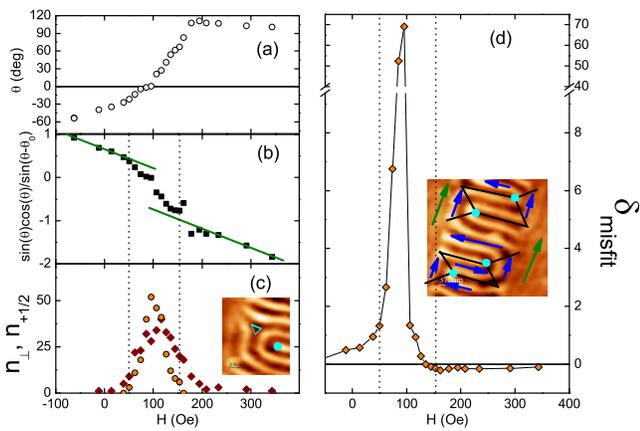}
 \end{center}
 \caption{(color online)
   (a) Field dependence of the orientation angle of $M_1$ relative to the lines.
    (b) Equilibrium condition for coherent rotation in the thin regions with $\theta_0 = 95^0$.
 Solid lines are linear fits to eq. (3). (c) Number
  of dislocations $n_\perp$ ($\diamond$) and $+1/2$ disclinations $n_{+1/2}$ ($\circ$) in the thick
  regions \textit{vs.} $H$. Dotted lines indicate the $H$ range of disclination observation.
 Inset is a detail of a disclination dipole of Burgers vector $3\lambda_2$.
 (d) $\delta_{misfit}$ \textit{vs.} $H$. Inset is
 a sketch of the in-plane magnetization configuration for
a buckled stripe pattern. $\circ$ and $\bigtriangleup$ indicate
$+1/2$ and $-1/2$ disclinations, respectively. }
 \end{figure}
Actually, it is in the intermediate range $50$ Oe $<H < 150$ Oe
where the most important structural transformations occur in the
magnetic stripe pattern of the thick lines: disclination dipoles
appear in the magnetic system and misfit strain reaches its
maximum (see Figs. 4(c)-(d)). These changes are a direct
consequence of the rotation of $M_1$ away from the field
direction. The stripe spacing projected along the interface
between thick and thin regions is
$\lambda_1^{||}=\lambda_1/\sin\theta$. Therefore, $\delta_{misfit}
= (\lambda_1^{||}-\lambda_2)/\lambda_2$ is gradually enhanced from
$\delta_{misfit} = 0.5$ to $1.3$ as $\theta$ goes from $-40^0$ at
remanence to $-23^0$ at 50 Oe. At the same time, a very large
density of misfit dislocations is observed in the MFM images.
Eventually, at $\delta_{misfit}\approx 1$, the distance between
simple dislocations of Burgers vector $b=\lambda_2$ reaches its
minimum value $D = \lambda_2$. Thus, in order to accommodate the
increasing strain, dislocations with larger $b = 2\lambda_2,
3\lambda_2,...$ should be nucleated. Instead, the MFM images
reveal the existence of a large number $n_{+1/2}$ of +1/2
disclinations within the thick regions in this intermediate field
range (see Fig. 4(c)).

In 2D, a dislocation is equivalent to a closely bound pair of +1/2
and -1/2 disclinations that can decay into a disclination dipole
either by the effect of temperature or strain
\cite{stripes1,copolymer}. One such disclination dipole of Burgers
vector $3\lambda_2$ is shown in the inset of Fig. 4(c). As
$\delta_{misfit}$ is dilative in the thick regions and compressive
in the thin ones, +1/2 disclinations are mostly observed in the
first case whereas -1/2 disclinations stay at the other side of
the interface. $n_{+1/2}$ reaches its maximum at the coercivity,
corresponding to the maximum misfit strain in the magnetic system.
Then, $n_{+1/2}$ decreases gradually with a certain lag relative
to the relaxation of $\delta_{misfit}$, until all the disclination
dipoles are recombinated into dislocations for $H$ above 150 Oe.

These +1/2 singularities in the stripe pattern are directly
coupled to the in-plane magnetization by the Bloch character of
the domain walls in between black-white stripes
\cite{supplemental}. In this framework, +1/2 disclinations are
equivalent to $180^0$ domain walls in $M_2$ together with a half
vortex closure structure, as sketched in the inset of Fig. 4(d).
In fact, domain walls in nanowires with in-plane magnetization
have already been described in terms of pairs of 1/2 topological
defects located at the nanowire edges \cite{walls}. The loss of
orientational order within the magnetic stripe pattern associated
to the presence of disclination dipoles is equivalent to the
nucleation of a multidomain structure during magnetization
reversal. This is seen from the comparison of Figs. 4(b) and (c):
the field range where +1/2 disclinations are observed corresponds
to the transition from negative to positive $H_D$ in the fits to
eq. (3) (i.e. negative to positive $M_2$). Also, it is interesting
to consider that, as in this field range $\theta$ approaches zero
(i.e. $M_2$ becomes perpendicular to $M_1$), the in-plane closure
domain structure associated to the disclination dipoles helps to
minimize the density of magnetic poles at the interfaces. Thus,
the magnetization reversal in this sample with deep nanostructured
grooves is a combination of two mechanisms: magnetization rotation
of $M_1$ due to the lines shape anisotropy and an incoherent
process that reverses $M_2$ by the nucleation of $180^0$ walls
linked to +1/2 disclinations within the magnetic stripe pattern
above a critical misfit strain $\delta_{misfit}=1$.

In summary, a lateral modulation of magnetic stripe periods has
been achieved by introducing a periodic thickness modulation in
PMA Nd-Co films. In the resulting lateral strained superlattice,
magnetic stripe patterns are controlled by the interplay between
interfacial misfit strain and shape anisotropy induced by
nanostructuration. For deep nanostructured grooves, high angle
boundaries appear in the 2D magnetic stripe pattern during
in-plane magnetization reversal. The structural changes in these
boundaries, driven by misfit strain, determine the magnetic
behavior of the system: the decay of high Burgers vector
dislocations into disclination dipoles above a critical misfit
strain can be directly linked to the nucleation of reversed
magnetic domains within the thicker regions.

Work supported by Spanish MICINN under grant FIS2008-06249.

\end{document}